\documentclass[twocolumn,pre]{revtex4}

\usepackage{dcolumn}
\usepackage{amsmath}

\usepackage{graphicx}
\usepackage{multirow}

\newcommand{\e}{\mathrm{e}}

\newcommand{\defn}{\textit}
\newcommand{\half}{\mbox{$\frac12$}}
\newcommand{\Ord}{\mathrm{O}}
\newcommand{\Aut}{\mathrm{Aut}}
\newcommand{\av}[1]{\langle#1\rangle}
\renewcommand{\vec}{\mathbf}
\newcommand{\mat}{\mathbf}

\newlength{\figurewidth}
\begin{document}

\title{Random graphs containing arbitrary distributions of subgraphs}
\author{Brian Karrer and M. E. J. Newman}
\affiliation{Department of Physics, University of Michigan, Ann Arbor, MI
  48109}
\affiliation{Santa Fe Institute, 1399 Hyde Park Road, Santa Fe, NM 87501}
\begin{abstract}
  Traditional random graph models of networks generate networks that are
  locally tree-like, meaning that all local neighborhoods take the form of
  trees.  In this respect such models are highly unrealistic, most real
  networks having strongly non-tree-like neighborhoods that contain short
  loops, cliques, or other biconnected subgraphs.  In this paper we propose
  and analyze a new class of random graph models that incorporates general
  subgraphs, allowing for non-tree-like neighborhoods while still remaining
  solvable for many fundamental network properties.  Among other things we
  give solutions for the size of the giant component, the position of the
  phase transition at which the giant component appears, and percolation
  properties for both site and bond percolation on networks generated by
  the model.
\end{abstract}

\maketitle

\section{Introduction}
\label{sec:intro}

It was pointed out by Rapoport in the 1940s~\cite{Rapoport48} and more
recently by Watts and Strogatz~\cite{WS98} that many observed networks
contain a statistically surprising number of triangles---sets of three
vertices connected by three edges.  More generally it has been noted that
many small subgraphs occur in networks in numbers greater than one would
expect purely on the basis of chance~\cite{Milo2002,Milo2004,BCC05}.  In
some cases there are good reasons for the frequent occurrence of a
particular subgraph.  Social networks, for instance, display a lot of
triangles because an individual's friends are likely themselves to be
friends.  In technological and biological networks certain subgraphs appear
to perform modular tasks that contribute to the networks' overall
operation~\cite{Milo2002} and hence may be evolutionarily favored.  In
other cases the reason for the appearance of the subgraphs is unclear, but
the empirical evidence for their presence is nonetheless convincing, and
hence if we wish to model these networks accurately the appropriate
subgraphs must be included.

Unfortunately, few practical models of complex networks exist that include
significant densities of arbitrary small subgraphs.  Random graph models
have been developed that incorporate numerous features of real-world
networks, including arbitrary degree distributions~\cite{MR95,NSW01},
correlations~\cite{PVV01,Newman02f}, bipartite and multi-partite structure,
hierarchy~\cite{CMN08}, vertex ordering~\cite{Karrer09a}, and even
geometry~\cite{Penrose03}, but until recently there were no equivalent
random graph models incorporating nontrivial densities of general
subgraphs.  Exponential random graphs do allow general subgraphs in
principle, but in practice they are difficult to analyze and moreover can
display pathological behaviors that limit their usefulness.

In this paper we introduce a class of models that build upon conventional
random graphs and incorporate nontrivial densities of arbitrary subgraphs,
while remaining exactly solvable in the limit of large network size for a
variety of properties both local and global, including degree
distributions, clustering coefficients, component sizes, percolation
properties, and others.  A number of new models have appeared in recent
years that achieve similar goals in other
contexts~\cite{Newman03e,SAS07,Bollobas2008,Newman09b,Gleeson2009,Gleeson20092,Miller2009},
and many of these are, as we will see, special cases of the general
formalism introduced in this paper.  We begin our discussion with a
description of a special case of the formalism, that of a network
containing only random edges and triangles, which was described previously
in Refs.~\cite{Newman09b,Miller2009}.  Then, using this model as a starting
point, we further develop the general foundations that will allow us to
model networks with any choice of subgraphs we please.

\section{A simple example}
\label{sec:edgetriangle}
Consider to begin with the standard Poisson random graph, famously studied
by Erd\H{o}s and R\'enyi in the 1950s and 60s~\cite{ER59,ER60}.  In this
model each distinct pair of vertices, out of~$n$ vertices total, is (or is
not) connected by an edge with independent probability $p=c/n$ (or $1-p$),
where $c$ is a constant.  A particular subgraph with $n_s$ vertices and
$m_s$ edges can occur in ${n\choose n_s}\simeq n^{n_s}/n_s!$ positions in
such a graph, each with probability
\begin{equation}
p^{m_s} (1-p)^{{n_s\choose 2}-m_s} \simeq \biggl( {c\over n} \biggr)^{m_s}.
\end{equation}
Thus the expected number of occurrences of such a subgraph is
$\Ord(n^{n_{s}-m_{s}})$.  If the subgraph is connected then $m_s\ge n_s-1$
with the equality applying only in the case of a tree.  Thus the number of
connected subgraphs grows with system size only for the case of trees and
for all other subgraphs is constant or decreasing, making the density of
all such non-tree subgraphs vanish in the large-$n$ limit.  A similar
argument applies to most of the extensions of the random graph.  These
graphs are thus ``locally tree-like''---all small connected subsets of
vertices within the network are trees.  This means, for instance, that such
graphs have a vanishing density of triangles in the limit of large size, in
strong contrast to real networks, which frequently have a very high
triangle density~\cite{note1}.

Although it is unrealistic, the locally tree-like nature of random graphs
is also the crucial feature that makes the model tractable.  The
calculations of the size of the giant component, the size distribution of
small components, the percolation features of the network, and many other
properties are all crucially dependent on the tree-like property.  This is
unfortunate, since the tree-like nature of the network is destroyed by the
introduction of a finite density of any subgraph that contains one or more
loops, which suggests that generalizations of random graph models
containing such subgraphs may be intrinsically unsolvable.  As we show in
this paper, however, this turns out not to be the case.  By exploiting
tree-like structure at a higher level, the level of the so-called ``factor
graph,'' we can introduce arbitrary distributions of subgraphs into the
network, including those containing loops, and still solve exactly for
global properties of the network, even though the network is now explicitly
not locally tree-like.  As a first demonstration of the process, consider
the following simple model, which we will call the ``edge--triangle''
model.

The edge--triangle model was proposed previously
in~\cite{Newman09b,Miller2009} as a way to incorporate the phenomenon of
clustering or transitivity into random graphs.  It generalizes the
configuration model, the standard model of a network with arbitrary degree
distribution~\cite{MR95,NSW01}.  In the configuration model one specifies
the number of edges attached to each vertex~$i$---the so-called ``degree
sequence''---as the fundamental parameters of the network.  In the
edge--triangle model one specifies instead the number~$t_i$ of triangles
that each vertex participates in along with the number~$s_i$ of ``single
edges,'' meaning edges that are not part of a triangle.  One can picture
vertex~$i$ as having $s_i$ ``stubs'' of edges emerging from it and $t_i$
corners of triangles.  Then an edge--triangle network is generated by
choosing stubs randomly in pairs and joining them to form complete edges
until no stubs remain, and also choosing triangle corners in threes and
joining them to form complete triangles until no corners remain.  The end
result is a network drawn uniformly at random from the set of all possible
matchings of the stubs and corners, and the edge--triangle model is defined
to be the ensemble of such matchings in which each appears with equal
probability.  Note that to create a complete matching we require that the
total number of stubs be a multiple of two and the total number of corners
a multiple of three.

The undirected networks generated by the edge--triangle model are clearly
not locally tree-like, since they contain triangles.  Yet one can still
proceed with analytic calculations.  Consider for instance the following
calculation of the size of the giant component (the extensive part of the
network in which any vertex can reach any other via at least one path).

Let us define the joint degree distribution $p(s,t)$ for our network to be
the fraction of vertices connected to $s$ single edges and $t$ triangles.
As with other random graph models, it's helpful to define a generating
function~$G_0$ for this degree distribution:
\begin{align}
G_0(z_1,z_2) = \sum_{st} p(s,t)\,z_1^s z_2^t.
\end{align}

Also important is the so-called ``excess degree distribution.''  Excess
degree is a property of the vertex one reaches by following an edge in a
network and is normally defined to be the number of edges connected to such
a vertex other than the edge one followed in the first place.  In the
edge--triangle model, where the ``degree'' of each vertex is denoted by the
two numbers $s$ and~$t$, there are now two corresponding excess degrees.
If one follows a single edge to reach a vertex then the excess degree is
given by the number~$t$ of triangles attached to that vertex and the
number~$s$ of single edges other than the edge via which we arrived.
Similarly if one follows a triangle to reach a vertex then the excess
degree is given by the number of single edges attached to that vertex and
the number of triangles other than the one via which we arrived.  It is
straightforward to show that the distributions of these excess degrees are,
respectively,
\begin{align}
q(s,t) &= \frac{(s+1)}{\av{s}} p(s+1,t), \\
r(s,t) &= \frac{(t+1)}{\av{t}} p(s,t+1),
\end{align}
where $\av{s}$ and $\av{t}$ are the average numbers of stubs and corners at
a vertex in the network as a whole.  The generating functions for the
excess degree distributions are
\begin{align}
G_1(z_1,z_2) &= \sum_{st} q(s,t)\,z_1^s z_2^t, \\
G_2(z_1,z_2) &= \sum_{st} r(s,t)\,z_1^s z_2^t.
\end{align}

The calculation of the giant component size now proceeds as follows.  Let
$u_1$ be the probability that the vertex reached by following a single edge
(an edge that is not part of a triangle) is not connected to the giant
component by any of its other edges or triangles, and let $u_2$ be the
probability that \emph{neither} of the vertices reached by following a
triangle is connected to the giant component by any of their other
triangles or edges.  If the vertex reached by following an edge is
connected to $s$ other edges and $t$ triangles then the probability that
none of them leads to the giant component is $u_1^s u_2^t$, where $s$ and
$t$ are distributed according to the excess degree distribution~$q(s,t)$.
Averaging over this distribution, we find that
\begin{equation}
u_1 = \sum_{st} q(s,t) u_1^s u_2^t = G_1(u_1,u_2).
\label{eq:tmu}
\end{equation} 
Similarly, if a vertex reached by following a triangle is is connected to
$t$ other triangles and $s$ edges then the probability that none of them
leads to the giant component is again $u_1^s u_2^t$, but with $s$ and~$t$
now distributed according to~$r(s,t)$.  Averaging over $r(s,t)$ then gives
an average probability of~$G_2(u_1,u_2)$ and the total probability for both
vertices reached via a triangle is the square of this quantity:
\begin{equation}
u_2 = \bigl[ G_2(u_1,u_2) \bigr]^2.
\label{eq:tmv}
\end{equation}

Finally, a randomly selected vertex with $s$ stubs and $t$ triangles is not
in the giant component if none of its neighbors are in the giant component,
which happens with probability $u_1^s u_2^t$ where $s$ and $t$ are
distributed according to~$p(s,t)$.  Averaging over $p(s,t)$, we find the
probability of not being in the giant component to be~$G_0(u_1,u_2)$, and
the probability~$S$ of being in the giant component is one minus this:
\begin{equation}
  S = 1 - G_0(u_1,u_2).
\label{eq:tms}
\end{equation}
Between them Eqs.~\eqref{eq:tmu} to~\eqref{eq:tms} give the size of the
giant component in our edge--triangle network as a fraction of the size of
the whole network.  While they cannot always be solved analytically, they
can be solved numerically by simple iteration: one makes an initial guess
about the values of~$u_1$ and~$u_2$ and iterates \eqref{eq:tmu}
and~\eqref{eq:tmv} to convergence, then substitutes the resulting values
into~\eqref{eq:tms}.  (The size of the giant component is only one example
of a quantity that can be calculated within the edge--triangle model.  For
other examples, including clustering coefficient and percolation properties
see Ref.~\cite{Newman09b}.)

The calculation of the giant component size in fact follows quite closely
the method used for other, locally tree-like random graph models such as
the configuration model~\cite{NSW01} and does not appear significantly more
complex despite the addition of triangles, which destroy the tree-like
property.  The reason for this is that, while the edge--triangle model is
indeed not tree-like in a naive sense, it is still tree-like at a higher
level, above the level of the triangles.  Specifically, in the limit of
large graph size, a finite-sized local neighborhood of a vertex in the
edge--triangle model is a connected graph whose largest biconnected
component (of which there can be many) is a triangle.  This means that if
we regard each triangle in the network as a single three-vertex unit (and
each single edge as a two-vertex unit) then local neighborhoods are
tree-like at the level of these units.  We will develop this idea further
shortly.

\section{A general model}
\label{sec:subgraphmodel}
The edge--triangle model provides a simple, solvable model of networks that
contain triangles.  It does, however, have some disadvantages.  In
particular, the probability that any two triangles connected to the same
vertex will share an edge vanishes in the limit of large graph size, a
direct consequence of the fact that the networks generated by the model are
tree-like above the level of triangles.  In real networks, by contrast, it
is a common occurrence that triangles share an edge, and hence the model is
unrealistic in this respect.

\begin{figure}
\begin{center}
\includegraphics[width=5cm]{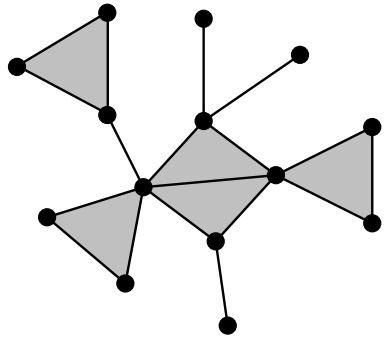}
\end{center}
\caption{A small network made of single edges, triangles, and ``diamond''
  subgraphs composed of two overlapping triangles.}
\label{fig:diamonds}
\end{figure}

One way to solve this problem would be to modify the model in some way to
encourage triangles to share edges, but this is unsatisfactory because, in
so doing, we would destroy the locally tree-like property of the network
(at the triangle level) that allows its solution.  Instead, therefore, we
propose an alternative approach: we consider two triangles that share an
edge to form a new network element, analogous to the triangle, but now with
four vertices instead of three---see Fig.~\ref{fig:diamonds}.  Our approach
is to create a model that introduces a specified distribution of these
larger elements in exactly the same way that we previously introduced
triangles.  More generally, it is possible to define a model in which we
introduce arbitrary distributions of any subgraphs we please.  (The
possibility of such a model was mentioned briefly in Refs.~\cite{Newman09b}
and~\cite{Miller2009}.)  As we will show, such models can always be viewed
as tree-like at a suitable higher level and thereby solved exactly for
properties both local and global in the limit of large size.

\subsection{Subgraphs and roles} 
\label{subsec:subgraphsets}
\begin{figure}
\begin{center}
\includegraphics[width=\columnwidth]{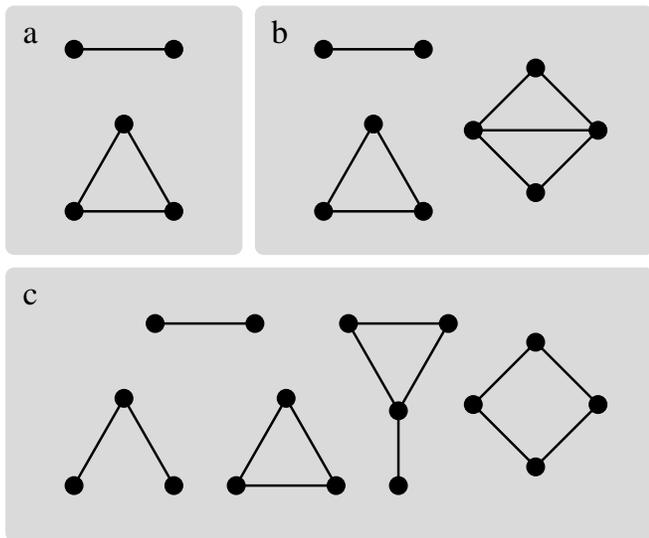}
\end{center}
\caption{Three examples of possible subgraph sets for the model described
  in the text.  (a)~The subgraph set for the edge--triangle model of
  Section~\ref{sec:edgetriangle}.  (b)~The subgraph set for the model with
  overlapping triangles shown in Fig.~\ref{fig:diamonds}.  (c)~A more
  extensive set that includes both singly and doubly connected subgraphs.}
\label{fig:sets}
\end{figure}

In the model we propose, we first specify a set of subgraphs that will be
added to the network.  Three examples of possible sets are shown in
Fig.~\ref{fig:sets}.  The network will be created by specifying the number
of each of the subgraphs attached to each vertex and then sampling randomly
from the (usually large) set of compatible networks.  The edge--triangle
model of Section~\ref{sec:edgetriangle} is an example of such a model in
which the set of subgraphs numbers just two---the single edge and the
triangle as shown in Fig.~\ref{fig:sets}b.  The model of this section
generalizes the edge--triangle model to arbitrary subgraph sets of
arbitrary size.

This generalization, however, introduces an important new feature to the
model that was not present in the edge--triangle model.  It is not
sufficient, in the general case, merely to specify how many copies of each
subgraph are connected to each vertex because the vertices in the subgraphs
can play more than one role.  Consider the diamond-shaped subgraph of
Fig.~\ref{fig:sets}b.  Two of the vertices in this subgraph have three
incident edges while the others have two.  Specifying only that a vertex
belongs to such a subgraph is therefore ambiguous.  We need to specify also
which role the vertex plays (a point made previously by
Miller~\cite{Miller2009}).  A vertex could, for example, participate in
five diamonds, playing the three-edge role in, say, four of them, and the
two edge role in the fifth.

\begin{figure}
\begin{center}
\includegraphics[width=6.5cm]{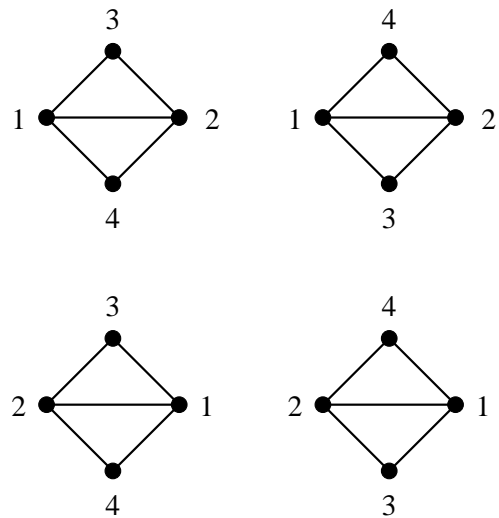}
\end{center}
\caption{Vertices 1 and 2 in this diamond-shaped subgraph constitute a role
  as defined in the text, as do vertices 3 and~4.}
\label{fig:orbit}
\end{figure}

The concept of a role in a subgraph can be made precise by employing
concepts from graph theory, specifically the concepts of automorphisms and
orbits.  Consider a subgraph in which each vertex has a unique identifying
label.  An \defn{automorphism} of the subgraph is a permutation of the
labels such that the set of label pairs joined by edges remains unchanged.
Consider Fig.~\ref{fig:orbit}, for instance, which shows the ``diamond''
subgraph from Fig.~\ref{fig:sets}b with integer labels on its vertices.
The four panels within the figure show four different automorphisms of the
graph so that there is, for instance always an edge between vertices 1
and~2 in each panel, regardless of the permutation of the labels.
Formally, the set of all automorphisms of a subgraph~$G$ forms a group,
which is called the automorphism group and denoted~$\Aut(G)$.

Now consider a particular vertex within our subgraph.  The \defn{orbit} of
the vertex is defined to be the set of other vertices with which it shares
at least one label under the label permutations of the automorphism group.
The left-most vertex in the subgraph of Fig.~\ref{fig:orbit}, for example,
shares labels 1 and 2 with the right-most vertex, but shares no labels with
either of the other vertices.  Hence the left and right vertices form an
orbit.  Similarly the top and bottom vertices form an orbit.  The orbits
correspond to the subgraph ``roles'' discussed above: they enumerate each
of the topologically distinct situations in which a vertex can find itself.

Unfortunately, the problem of calculating the orbits of the automorphism
group of a graph is computationally hard and no polynomial-time algorithm
is known.  Our focus here, however, is on relatively small subgraphs for
which roles can be found quite simply by hand.

Once subgraph roles are defined then we can specify completely the number
of subgraphs in which each vertex participates and the manner in which it
does so.  We assign to each vertex a set of role indices~$d_1,d_2,\ldots$,
specifying the number of times each vertex plays each possible role.  For
instance, in the subgraph set shown in Fig.~\ref{fig:sets}b, there are four
different roles---one for each of the first two subgraphs and two for the
third subgraph.  Thus each vertex would in this case be assigned four role
indices $d_1\ldots d_4$ giving the number of times it plays each of these
roles.  The set of role indices for a given vertex form a vector~$\vec{d}$
with~$c$ non-negative integer elements, where $c$ is number of different
roles in the subgraph set.  This vector plays a part similar to that played
by the degree in conventional network models and, by analogy with the
degree sequence in such models, we will call the set of vectors for all
vertices the \defn{role sequence}.

Just as with degree sequences in other models, not all role sequences
correspond to possible networks.  Consider again the diamond subgraph of
Fig.~\ref{fig:orbit}, with its two roles.  Because both roles appear two
times in this subgraph, the total number of times vertices in a network
participate in each of the roles must be equal to the same multiple of two.
More generally, the total number of times vertices play a particular role
in a particular subgraph must always be an multiple of the number of times
that role appears in the subgraph, and moreover that multiple must be the
same for all roles of the given subgraph.  Role sequences that fulfill
these conditions are said to be \defn{graphical}.

\subsection{Network creation}
\label{subsec:buildnets}
Given a graphical role sequence, a random network drawn from the ensemble
of our proposed model can be generated in a straightforward fashion.  The
role indices for each vertex dictate the numbers of stubs of each type
attached to the vertex and creation of the network involves working through
each of the subgraph types in turn and repeatedly choosing a set of stubs
at random in the appropriate combination for that subgraph and connecting
them together to make the subgraph in question.  When all stubs for a given
subgraph type have been exhausted we move on to the next subgraph until the
list of subgraphs is also exhausted.  The end result is a matching of stubs
to form subgraphs, drawn uniformly from the set of all possible such
matchings.

It is possible in the creation of a given subgraph that two or more stubs
attached to the same vertex will be drawn from the set of available stubs.
If this happens the resulting network will contain either an edge that
connects a vertex to itself---a self-edge---or multiple edges between the
same pair of vertices---a multiedge---or both.  In the model we propose,
such edges are allowed to exist, even though they are prohibited in many
real-world networks.  A similar situation arises in the standard
configuration model of graph theory and, as in that model, the densities of
self-edges and multiedges in our model both vanish in the limit of large
network size as~$1/n$ and hence can be neglected in this limit.  It is
possible to create models that generate ``simple graphs,'' i.e.,~graphs
without self-edges or multiedges, but such models are considerably more
complicated to work with, both analytically and computationally.

In the calculations presented here, we will not assume that we are given a
complete role sequence for all vertices, but only that we are given a role
distribution~$p(\vec{d})$, which specifies the probability that a randomly
chosen vertex has role vector~$\vec{d}$ (or equivalently specifies the
fraction of vertices having role vector~$\vec{d}$ in the large-$n$ limit).
The network itself will then be defined by drawing a role sequence from
this distribution and forming a random matching of the set of stubs
specified by the sequence.

The constraint that the role sequence must be graphical imposes a
corresponding constraint on the role distribution.  Let $\av{d_r} =
\sum_{\vec{d}} d_r p(\vec{d})$, where the sum is over all possible values
of the role vector~$\vec{d}$.  Then the expected number of role-$r$ stubs
in the entire network is $n\av{d_r}$ and the number of occurrences of the
corresponding subgraph is $n\av{d_r}/n_r$, where $n_r$ is the number of
times that role~$r$ appears in a single instance of the subgraph.  We can
calculate a corresponding figure for the number of occurrences of the
subgraph using any of its other roles and all of these figures must be
equal if the role sequence is to be graphical.  Thus we must have
\begin{equation}
\frac{\av{d_r}}{n_r} = \frac{\av{d_s}}{n_s}
\end{equation}
for all pairs~$r,s$ of roles within the same subgraph.

Assuming we have a role distribution~$p(\vec{d})$ that satisfies this
constraint, the procedure for generating a network is first to draw a
complete role sequence $\vec{d}_i$, $i=1\ldots n$ for the network.  It is
possible that, despite the constraint on~$p(\vec{d})$, this sequence will
not be graphical, because of statistical fluctuations in the role vectors
drawn.  If this is the case, we choose a vertex uniformly at random,
discard its role vector and draw another from~$p(\vec{d})$.  We repeat this
process until the role sequence is graphical.  Then the network itself is
generated by random matching of stubs as described above.

The role distribution also fixes the conventional degree distribution for
the network.  Each role $r$ has some number $k_r$ of associated edges, so a
vertex with role indices $d_1\ldots d_c$ has $\sum_r k_r d_r$ edges.  The
fraction of vertices with total degree~$k$ is then
\begin{equation}
p(k) = \sum_{\vec{d}} p(\vec{d}) \delta\Bigl(k,\sum_r k_r d_r\Bigr).
\label{eq:degdist}
\end{equation}
While nothing in the above strictly demands it, we will here consider only
\emph{sparse} graphs for which the average total degree of a
vertex~$\av{k}$ is constant in the limit of large~$n$.  From
Eq.~\eqref{eq:degdist} we have
\begin{align}
\av{k} &= \sum_k k p(k)
  = \sum_k k\sum_{\vec{d}} p(\vec{d}) \delta\Bigl(k,\sum_r k_r d_r\Bigr)
    \nonumber\\
 &= \sum_r k_r \sum_{\vec{d}} d_r p(\vec{d})
  = \sum_r k_r \av{d_r}.
\label{eq:avk}
\end{align}

\section{Bipartite graph representation}
\label{sec:bipartitepic}
Like the edge-triangle model (of which it is a superset), this subgraph
model is not in general locally tree-like but can be thought of as
tree-like at a higher level.  If one considers the subgraphs as coherent
graph units in their own right, then the network is still tree-like at the
level of these units.  This idea can be made more precise as follows.

Our subgraph model has an alternative representation as a bipartite graph:
a network with two types of vertices and edges running only between unlike
types.  In this representation one of the types of vertices corresponds to
the vertices of the original network while the other corresponds to the
subgraphs, and each original vertex is connected by an edge to the
subgraphs in which it participates.  (In other circumstances this
representation would be called a \defn{factor graph}.)  In order to
distinguish the different roles that a vertex can play in a subgraph the
edges can be labeled with the appropriate role numbers~$r$.

A vertex in the original graph having role vector
$\vec{d}=(d_1,d_2,\ldots,d_c)$ is represented in the bipartite graph by a
vertex with $d_1$ incident edges marked with role label~$1$, $d_2$~edges
marked with label~2, and so forth.  On the other side of the bipartite
graph, every vertex representing a given subgraph in the original network
has the same number of stubs, again labeled by role, one for each vertex in
the subgraph.

\begin{figure}
\begin{center}
  \includegraphics[width=\columnwidth]{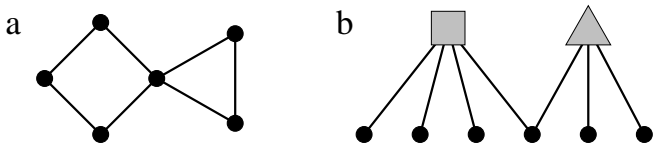}
\end{center}
\caption{(a)~A small network composed of a square subgraph and a triangle
  and (b)~the bipartite representation of the same network.}
\label{fig:bipartite}
\end{figure}

The process of creating a network in the original representation of the
model is equivalent in the bipartite representation to a random matching of
stubs subject to the constraint that every edge created joins a single
vertex stub to a single subgraph stub having the same role label.
Figure~\ref{fig:bipartite} illustrates the equivalence between the two
representations.

The important point to notice about this construction is that in the limit
of large graph size the bipartite graph is strictly locally tree-like for
the same reason that standard random graphs are locally tree-like---the
density of loops of any fixed size falls off as $1/n$ as $n$ becomes large.
Thus computations can be carried out on the bipartite graph using standard
techniques for such networks~\cite{NSW01} and the results can then be
translated back into the language of the original network to calculate
properties of the subgraph model.  In the following sections we give a
number of such calculations of increasing complexity.  The first and
simplest is the calculation of the expected number of occurrences of each
subgraph in the subgraph model.

\subsection{Subgraph counts}
\label{subsec:subgraphcount}
If we want to know how many times a particular subgraph occurs in our
subgraph model we must allow for three distinct ways in which a subgraph
can be created: it can be added explicitly as one of the set of allowed
subgraphs; it can be added implicitly as part of a larger subgraph; or it
can be created by putting two or more subgraphs together.  For instance, a
triangle can be formed if we explicitly add a triangle to the network, if
we add a larger subgraph containing a triangle (such as the diamond
subgraph in Fig.~\ref{fig:sets}b), or we can add three independent single
edges to the network that happen to coincide and create a triangle.

There is an important distinction to be drawn here between subgraphs that
are biconnected and those that are only singly connected.  Recall that a
biconnected graph is one in which there exist at least two
vertex-independent paths between any pair of vertices, or equivalently in
which at least two vertices must be removed to disconnect any pair of
vertices.  It is trivial to see that a biconnected subgraph can only be
created by joining two or more other subgraphs if the joined subgraphs meet
at at least two places---if they meet at only one then the removal of that
one vertex will disconnected them and hence they cannot have been
biconnected.  The locally tree-like property of the bipartite
representation of the subgraph model, however, implies that the probability
of two subgraphs of given size meeting at two places vanishes in the limit
of large graph size.  Such subgraphs would form a loop of length four in
the bipartite graph composed of the two subgraphs and the two vertices at
which they meet.  Since the bipartite graph is locally tree-like such loops
do not occur (or more properly occur with vanishing density in the limit of
large~$n$).  This implies that the density of biconnected subgraphs formed
by the joining of others vanishes in our subgraph model and hence that if
we wish to know about biconnected subgraphs we need concern ourselves only
with those added either explicitly to the network or implicitly as part of
a larger subgraph.  And the expected number of such subgraphs is trivial to
calculate---we simply sum up the appropriate numbers.

One important consequence of this result is that we can calculate the
clustering coefficient of our network in a straightforward manner.  The
clustering coefficient is defined to be~\cite{NSW01}
\begin{equation}
C = \frac{3\times\mbox{(number of triangles in network)}}%
  {\mbox{(number of connected triples)}} = \frac{3N_\Delta}{N_3}.
\label{eq:clustering}
\end{equation}
Since the triangle is a biconnected subgraph (in fact the smallest such
subgraph), the number of triangles $N_{\Delta}$ is given simply by counting
how many triangles appear in each allowed subgraph, multiplying by the
expected number of the respective subgraphs, and summing over subgraphs.
The number of connected triples~$N_3$ is given by
\begin{equation}
n\sum_k {k\choose2} p(k)
  = \half n\biggl\langle \biggl[ \sum_r k_r d_r \biggr]^2
    - \sum_r k_r d_r \biggr\rangle,
\end{equation}
and hence we can evaluate Eq.~\eqref{eq:clustering}.

For singly connected subgraphs the situation is more complicated.  Such
subgraphs can be created by joining together two or more subgraphs at
single vertices, which happens frequently even in locally tree-like
networks.  As a result it is not trivial to calculate the expected number
of singly connected subgraphs in a network, except in a few special cases.
(The connected triple or 2-star of Eq.~\eqref{eq:clustering} is an example
of a singly connected subgraph whose number can be calculated in a
straightforward fashion, but this is the exception rather than the rule.)

\subsection{Giant component}
\label{subsec:giantcomponent}
One can also exploit the bipartite representation of the subgraph model to
find the size of the giant component in the model.  As with the treatment
of the edge--triangle model in Section~\ref{sec:edgetriangle}, we introduce
excess degree distributions, but defined now in terms of the bipartite
network.  Consider a subgraph node in the bipartite network and imagine
following one of the edges connected to it, an edge labeled with role~$r$,
to the vertex at its other end.  Then, if that vertex has overall role
indices $d_1,\ldots,d_r+1,\ldots,d_c$ and we exclude the edge along which
we arrived, it will have ``excess'' role indices
$d_1,\ldots,d_r,\ldots,d_c$.  Moreover, since the probability of our
arriving at a vertex with role index~$d_r$ is proportional to~$d_r$ the
excess role indices are distributed according to the probability
distribution
\begin{equation}
q_r(\vec{d}) = {d_r+1\over\av{d_r}}\,p(d_1,\ldots,d_r+1,\ldots,d_c),
\end{equation}
and there is a corresponding distribution for every other role.

As in the edge--triangle model it is convenient to keep track of the
various role distributions using their generating functions, defined by
\begin{align}
\label{eq:defsg0}
G_0(\vec{z}) = \sum_{\vec{d}} p(\vec{d}) z_1^{d_1}\ldots z_c^{d_c}, \\
G_r(\vec{z}) = \sum_{\vec{d}} q_r(\vec{d}) z_1^{d_1}\ldots z_c^{d_c}.
\label{eq:defsgr}
\end{align}
Note that given the generating function~$G_0(\vec{z})$ we can calculate the
generating function~$H(z)$ for the conventional degree distribution in a
straightforward manner using Eq.~\eqref{eq:degdist} thus:
\begin{align}
H(z) &= \sum_k p(k)\,z^k
  = \sum_k \sum_{\vec{d}} p(\vec{d})\,\delta\Bigl(k,\sum_r k_r d_r\Bigr) z^k
    \nonumber\\
 &= \sum_{\vec{d}} p(\vec{d}) z^{k_1d_1}\ldots z^{k_cd_c} \nonumber\\
 &= G_0\bigl(z^{k_1},\ldots,z^{k_c}\bigr),
\label{eq:gfdegree}
\end{align}
where $k_r$ is, as previously, the number of edges a vertex gains by virtue
of playing role~$r$ in a subgraph.

We also need to define generating functions for the role distributions on
the other side of the bipartite graph, the side that represents the
subgraphs.  These generating functions, however, take a relatively simple
form, since every role belongs to only one subgraph and every instance of a
given subgraph contains the same distribution of roles.  Let us define
$n_r$ as before to be the number of times role~$r$ occurs in its own
subgraph, and let us give the subgraphs unique labels such that $g_r$ is
the label of the graph in which role~$r$ appears.  Then the generating
function for the excess role distribution of a subgraph node reached by
following an edge representing role~$r$ is
\begin{equation}
F_r(\vec{z}) = {1\over z_r} \prod_{s=1}^c z_s^{n_s \delta_{g_rg_s}},
\label{eq:defsfrz}
\end{equation}
where $\delta_{ij}$ is the Kronecker delta.  (We can in principle write
down a generating function for the normal (non-excess) role distribution of
the subgraph nodes, but we omit it because it's not needed for our
calculations.)

Armed with these generating functions, we can compute the size of the giant
component in the network as follows~\cite{note2}.  Define $u_r$ to be the
probability that a vertex is \emph{not} connected to the giant component as
a result of its playing role~$r$ in a given subgraph.  This occurs if and
only if none of the other vertices in that subgraph, regardless of role,
are themselves members of the giant component, or, in the language of the
bipartite representation, if none of the other edges connected to the
appropriate subgraph node lead to vertices that are in the giant component.

Similarly, let $v_r$ be the probability that a vertex that plays role~$r$
in a particular subgraph is not a member of the giant component by virtue
of any of the \emph{other} roles it plays.  In the language of the
bipartite representation, none of the other edges incident on such a vertex
connect it to the giant component.

In terms of these variables we have
\begin{align}
u_r = {1\over v_r} \prod_{s=1}^c v_s^{n_s \delta_{g_rg_s}}
    = F_r(v_1,\ldots,v_c), \\
v_r = \sum_{\vec{d}} q_r(\vec{d}) \prod_{s=1}^c u_s^{d_s}
    = G_r(u_1,\ldots,u_c),
\end{align}
or, in vector notation, $\vec{u} = \vec{F}(\vec{v})$ and $\vec{v} =
\vec{G}(\vec{u})$.  Being primarily concerned with~$\vec{u}$, we can also
eliminate $\vec{v}$ and write $\vec{u}$ as the solution of the fixed point
equation
\begin{equation}
\vec{u} = \vec{F}(\vec{G}(\vec{u})).
\label{eq:u}
\end{equation}
Then the probability that a vertex with role indices $d_1\ldots d_c$ is not
in the giant component is $\prod_r u_r^{d_r}$ with $d_r$ distributed
according to~$p(\vec{d})$, so that the average probability is
\begin{equation}
\sum_{\vec{d}} p(\vec{d}) \prod_r u_r^{d_r} = G_0(\vec{u}),
\end{equation}
and the average probability that a vertex \emph{is} in the giant component
is one minus this quantity:
\begin{equation}
S = 1 - G_0(\vec{u}).
\label{eq:s}
\end{equation}
Between them, Eqs.~\eqref{eq:u} and~\eqref{eq:s} give us a
complete solution for the size of the giant component.  As with the
edge--triangle model the equations are often not solvable in closed form,
but can be solved numerically by simple iteration starting from a suitable
initial value for~$\vec{u}$.

As a first example, consider again the edge--triangle model, for which
there are two subgraphs, the edge and the triangle, with one role each and
generating functions
\begin{equation}
F_1(z_1,z_2) = z_1,\qquad F_2(z_1,z_2) = z_2^2.
\end{equation}
The excess degree generating functions for the vertices are
\begin{align}
G_1(z_1,z_2) &= {1\over\av{d_1}} \sum_{\vec{d}} (d_1+1) p(d_1+1,d_2)
                                z_1^{d_1} z_2^{d_2}, \\
G_2(z_1,z_2) &= {1\over\av{d_2}} \sum_{\vec{d}} (d_2+1) p(d_1,d_2+1)
                                z_1^{d_1} z_2^{d_2},
\end{align}
and Eqs.~\eqref{eq:u} and~\eqref{eq:s} reduce to the two
equations
\begin{equation}
u_1 = G_1(u_1,u_2),\qquad
u_2 = \bigl[ G_2(u_1,u_2) \bigr]^2,
\end{equation}
which are identical to Eqs.~\eqref{eq:tmu} and~\eqref{eq:tmv}.  And the
size of the giant component as a fraction of the size of the whole network
is given by $S = 1 - G_0(u_1,u_2)$ as before.

As a more complicated example consider a network built from the subgraphs
shown in Fig.~\ref{fig:sets}b: single edges, triangles, and diamonds.  Of
the four roles let us label the ends of single edges role~1, the corners of
the triangles role~2, and the two roles in the diamond roles 3 and~4 (which
is which will not matter).  Now consider the role distribution
$p(\vec{d})=p_1(d_1)p_2(d_2)p_{34}(d_3,d_4)$ where
\begin{align}
p_1(d_1) &= \e^{-c_1} {c_1^{d_1}\over {d_1}!}, \qquad
p_2(d_2) = \e^{-c_2} {c_2^{d_2}\over {d_2}!}, \\
p_{34}(d_3,d_4) &= (1-2a) \delta_{r_3,0}\delta_{r_4,0} \nonumber\\
                   & \qquad {} + a [ \delta_{r_3,0}\delta_{r_4,1}
                        + \delta_{r_3,1}\delta_{r_4,0} ].
\label{eq:defsa}
\end{align}
In other words, participation is Poisson distributed with means $c_1$ and
$c_2$ for roles 1 and~2, and vertices participate either in one diamond
with equal probability~$a$ of taking role 3 or~4, or in no diamonds with
probability~$1-2a$.

\begin{figure}
\begin{center}
\includegraphics[width=8cm]{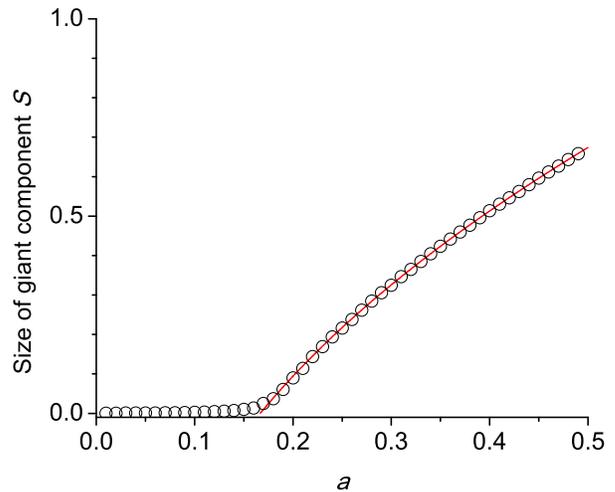}
\end{center}
\caption{The size~$S$ of the giant component in the network of edges,
  triangles, and diamonds described in the text, with the average role
  indices for edges and triangles fixed at $\av{d_1}=\frac14$ and
  $\av{d_2}=\frac18$, plotted as a function of the parameter~$a$ that
  controls the two role indices for the diamonds (see
  Eq.~\eqref{eq:defsa}).  The solid line represents the analytic result and
  the circles are simulation results averaged over 100 networks with $10^5$
  vertices each.}
\label{fig:graph1}
\end{figure}

This particular distribution is chosen because it has a nontrivial but
still relatively simple solution: after some work it can be shown that the
size~$S$ of the giant component obeys
\begin{equation}
S = 1 - (1-2a) \e^{-c_1 S-c_2 S(2-S)} - 2a \e^{-4c_1 S-4c_2 S(2-S)}.
\end{equation}
We show the form of this solution as a function for~$a$ for one particular
choice of parameters in Fig.~\ref{fig:graph1}.

\subsection{Position of the phase transition}
As with other random graph models, the fixed point equation,
Eq.~\eqref{eq:u}, has a trivial solution at $u_r=1$ for all~$r$
corresponding to the state in which there is no giant component in the
network and this fixed point undergoes a transcritical bifurcation at the
point at which a giant component appears.  We can locate the bifurcation,
and hence the appearance of the giant component, by linear stability
analysis of the fixed point.  We write $u_r=1-\epsilon_r$ and expand to
first order in the small quantities $\epsilon_r$, which gives
$\boldsymbol{\epsilon} = \mat{M}\boldsymbol{\epsilon}$ where $\mat{M}$ is
the $c\times c$ matrix with elements
\begin{align}
M_{rs} &= \sum_t \bigl( n_t \delta_{g_rg_t} - \delta_{rt} \bigr)
          {\partial G_t\over\partial z_s} \biggl|_{\vec{z}=\vec{1}}.
\label{eq:m}
\end{align}
Using the definition of $G_r(\vec{z})$ given in Eq.~\eqref{eq:defsgr} we
can show that
\begin{equation}
{\partial G_t\over\partial z_s} \biggl|_{\vec{z}=\vec{1}}
  = {\av{d_sd_t}\over\av{d_t}} - \delta_{st},
\end{equation}
and hence that
\begin{equation}
M_{rs} = \delta_{rs} - {\av{d_rd_s}\over\av{d_r}} - n_s \delta_{g_rg_s}
         + \sum_t {\av{d_sd_t}\over\av{d_t}} n_t \delta_{g_rg_t}.
\end{equation}

Physically the matrix element $M_{rs}$ measures a branching ratio for the
locally tree-like bipartite graph: a vertex that plays role~$r$ shares the
relevant subgraph with some number of other vertices and $M_{rs}$ measures
the average number of times those other vertices collectively play role~$s$
in further subgraphs.

Now consider a set of randomly chosen vertices in a large network and
suppose we grow that set repeatedly by adding to it all vertices with which
its members share a subgraph.  If we focus on the boundary of the
set---meaning those vertices added on the most recent step---and represent
the number of times the boundary vertices play roles 1 to~$c$ by a
$c$-component vector, then the expected value of the vector is multiplied
on each step by one factor of~$\mat{M}$.  If the sum of the vector elements
grows we have a giant component and if it dwindles to zero, so that the set
stops growing, then we do not, meaning that we have a giant component if
and only if at least one eigenvalue of~$\mat{M}$ is greater than one.  If
all eigenvalues are less than one then there is no giant component, and if
one or more eigenvalues are exactly equal to one and none are greater then
we are precisely at the phase transition point at which the giant component
appears.

In general the eigenvalues of~$\mat{M}$ are not trivial to find, but in
some cases the problem simplifies.  Consider, for instance, the case in
which the role indices~$d_r$ are independent and Poisson distributed, in
which case $\av{d_rd_s}=\av{d_r}\av{d_s} + \av{d_r}\delta_{rs}$ so that
\begin{equation}
M_{rs} = (N_r-1) \av{d_s},
\end{equation}
where $N_r = \sum_t n_t \delta_{g_rg_t}$ is the number of vertices in the
subgraph in which role~$r$ appears.  As an outer product of two vectors,
this matrix is defective, having only a single eigenvector with
corresponding eigenvalue $\lambda=\sum_r (N_r-1)\av{d_r}$.  Thus in this
network a giant component exists if and only if
\begin{equation}
\sum_r (N_r-1)\av{d_r} > 1.
\end{equation}
In simple language, this equation says that a giant component exist if and
only if the average number of vertices with which a random vertex shares a
subgraph is greater than one.

Note that the degree~$k_r$ of role~$r$ within its subgraph trivially is
never greater than $N_r-1$ and hence when we are precisely at the
transition point at which the giant component forms the average degree,
Eq.~\eqref{eq:avk}, satisfies
\begin{equation}
\av{k} = \sum_r k_r \av{d_r} \le \sum_r (N_r-1)\av{d_r} = 1.
\label{eq:comper}
\end{equation}
Recall that in the standard random graph of Erd\H{o}s and R\'enyi the
transition occurs precisely at $\av{k}=1$ and Eq.~\eqref{eq:comper} thus
tells us, in some sense, that the transition happens ``earlier'' in this
subgraph model, or at least no later---in general a giant component will
form when the average degree is less than one.  The physical insight behind
this observation is that belonging to a subgraph guarantees a vertex
connections to all the other vertices in that subgraph, which vertices may
be significantly greater in number than merely the immediate neighbors of
the first vertex.

Another simple solvable case is that of a network composed of two types of
subgraph, each with a single role.  The edge--triangle model of
Section~\ref{sec:edgetriangle} is a special case of such a model, but the
general case is also solvable.  The matrix~$\mat{M}$ takes the form
\begin{equation}
\mat{M} = \begin{pmatrix}
  (n_1-1) \dfrac{\av{d_1^2}-\av{d_1}}{\av{d_1}} &
  (n_1-1) \dfrac{\av{d_1d_2}}{\av{d_1}} \\
  \\
  (n_2-1) \dfrac{\av{d_1d_2}}{\av{d_2}} &
  (n_2-1) \dfrac{\av{d_2^2}-\av{d_2}}{\av{d_2}}
          \end{pmatrix},
\end{equation}
and the largest eigenvalue is $\lambda=\half(\tau+\sqrt{\tau^2-4\Delta})$
where $\tau$ and $\Delta$ are the trace and determinant of the matrix
respectively.  There is a giant component if and only if this eigenvalue is
greater than one, or equivalently if $\sqrt{\tau^2-4\Delta}>2-\tau$.  This
condition is satisfied if either $\tau>2$ or $\tau>\Delta+1$.  In terms of
the matrix elements, these inequalities can be written
\begin{equation}
(n_1-1) \frac{\av{d_1^2}-\av{d_1}}{\av{d_1}}
  + (n_2-1) \frac{\av{d_2^2}-\av{d_2}}{\av{d_2}} > 2,
\label{eq:condition1}
\end{equation}
and
\begin{equation}
\frac{\av{d_1d_2}^2}{\av{d_1}\av{d_2}} - 
\biggl[ \frac{\av{d_1^2}}{\av{d_1}} - {n_1\over n_1-1} \biggr]
\biggl[ \frac{\av{d_2^2}}{\av{d_2}} - {n_2\over n_2-1} \biggr] > 0.
\label{eq:condition2}
\end{equation}
Note that Eq.~\eqref{eq:condition1} does not depend on the correlation term
$\av{d_1d_2}$ between the two role indices: in physical terms it says
simply that there is a giant component if the densities of the two
subgraphs independently are enough to create one.  A giant component can,
however, also be created by correlations between the placement of the
subgraphs even when the overall density of subgraphs would otherwise be
inadequate, and giant components of this kind are described by
Eq.~\eqref{eq:condition2}.  Thus, for instance, if the vertices with many
of one subgraph also have many of the other, then these high-degree
vertices may join up to form a giant component even if there would be none
were the same subgraphs allocated to different vertices.

In the examples we have examined, the size of the giant component and the
position of the phase transition are determined solely by the role
distribution~$p(\vec{d})$ and the numbers~$n_r$ of roles in their
respective subgraphs.  The actual shape of the subgraphs themselves does
not matter---once one vertex in a subgraph is in the giant component, the
rest automatically are as well, regardless of patterns of connection within
the subgraph.  Thus, for instance, a network composed of a given
distribution of cliques of given sizes would have a giant component of the
same size as a network composed of the same distribution of loops of the
same sizes.  Not all network properties, however, show this behavior.  In
the next section we look at percolation on our networks, for which the
shapes of the subgraphs matter greatly.

\subsection{Percolation}
\label{subsec:percolation}
Site and bond percolation have important implications for network
resilience~\cite{CEBH00,CNSW00} and the behavior of spreading processes on
networks~\cite{Mollison77,Grassberger82,Newman02c}.  In site percolation,
each vertex of a graph is present, functional, or ``occupied'' with
independent probability~$\phi_s$.  In bond percolation each edge is
similarly occupied with independent probability~$\phi_b$.

It turns out that percolation on networks generated by our subgraph model
can be treated using the same machinery developed above for calculating
properties of the giant component.  The only thing that needs to change is
the generating function~$F_r(z)$, Eq.~\eqref{eq:defsfrz}, for the
probability distribution of the number of vertices of each role that a
vertex of role~$r$ can reach in its subgraph.  In our previous treatment
this distribution took a particularly simple form, since a vertex of
role~$r$ could by definition reach all other vertices in its subgraph.
Once we introduce percolation, however, some vertices in the subgraph may
become unreachable, because there is no path of occupied vertices or edges
along which to travel.

We will here consider the general case of simultaneous site and bond
percolation---both vertices and edges may be occupied (or not) with
probabilities $\phi_s$ and $\phi_b$ (or $1-\phi_s$ and $1-\phi_b$).
Percolation on vertices or edges alone, i.e.,~conventional site or bond
percolation, is the special case of this more general process when either
$\phi_s$ or $\phi_b$ equals one.

Let us denote by $p_r(\phi_s,\phi_b; \vec{d})$ the probability that an
occupied vertex of role~$r$ in a subgraph is connected via occupied
vertices and edges to $d_1\ldots d_c$ occupied vertices of roles $1\ldots
c$ in the same subgraph, given the values $\phi_s$ and~$\phi_b$ of the
occupation probabilities.  Then our generating function $F_r$ is defined by
\begin{equation}
F_r(\phi_s,\phi_b;\vec{z}) = \sum_\vec{d} p_r(\phi_s,\phi_b; \vec{d})
                             \prod_{s=1}^c z_s^{d_s}.
\end{equation}
In terms of this function, the size of the giant percolation cluster is
given by
\begin{equation}
\vec{u} = \vec{F}(\vec{G}(\vec{u})),\qquad S = \phi_s[1-G_0(\vec{u})].
\end{equation}
(See Eqs.~\eqref{eq:u} and~\eqref{eq:s} for the corresponding equations in
our earlier calculation.)  The additional factor of $\phi_s$ in the second
equation accounts for the fact that a vertex can only be in the giant
component if it is itself occupied.  All we need now to complete the
calculation is the form of~$F_r$.

\begin{table*}[t]
\setlength{\tabcolsep}{10pt}
\centering\normalsize
\begin{tabular}{c|c|l}
Subgraph & Role & Generating function~$F_r$ \\
\hline
& & \\
\begin{minipage}[c]{1cm}\includegraphics[width=1cm]{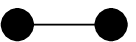}\end{minipage} &
\begin{minipage}[c]{1cm}\includegraphics[width=0.26cm]{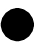}
  \end{minipage} &
$1-\phi_s \phi_b + \phi_s \phi_b z$ \\ 
& & \\
\hline
& & \\
\begin{minipage}[c]{1cm}\includegraphics[width=1cm]{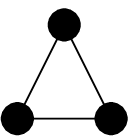}
  \end{minipage} &
\begin{minipage}[c]{1cm}\includegraphics[width=0.26cm]{FilledDot.eps}
  \end{minipage} &
$(1-\phi_s \phi_b)^2 + 2\phi_s \phi_b(1-\phi_s \phi_b(2-\phi_b))z +
  \phi_s^2 \phi_b^2(3-2\phi_b)z^2$ \\
& & \\
\hline
& & \\
\begin{minipage}[c]{1cm}\includegraphics[width=1cm]{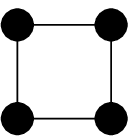}\end{minipage} &
\begin{minipage}[c]{1cm}\includegraphics[width=0.26cm]{FilledDot.eps}
  \end{minipage} &
$(1-\phi_s\phi_b)^2 + 2\phi_s\phi_b(1-\phi_s\phi_b)^2z +
3\phi_s^2\phi_b^2(1-\phi_s\phi_b(2-\phi_b))z^2 +
\phi_s^3\phi_b^3(4-3\phi_b)z^3$ \\
& & \\
\hline
& & \\
\multirow{2}{*}{\begin{minipage}[c]{1cm}
\null\vspace{4mm}\includegraphics[width=1cm]{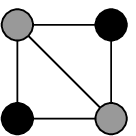}\end{minipage}} &
\begin{minipage}[c]{1cm}
\begin{center} 1 \\ \includegraphics[width=0.26cm]{FilledDot.eps}\end{center}
\end{minipage} &
\begin{minipage}[c]{13cm}
\raggedright
$(1-\phi_s \phi_b)^2 + 2\phi_s \phi_b(1-\phi_s\phi_b)(1-\phi_s
\phi_b(2-\phi_b))z_2+\phi_s^2 \phi_b^2(3-2\phi_b)(1-\phi_s
\phi_b(2-\phi_b))z_2^2 + 2\phi_s^2\phi_b^2( 1 - 3\phi_s\phi_b +
3\phi_s\phi_b^2 - \phi_s\phi_b^3) +\phi_s^3
\phi_b^3(8-11\phi_b+4\phi_b^2)z_1z_2^2$
\end{minipage} \\
& & \\
& \begin{minipage}[c]{1cm}
\begin{center} 2 \\ \includegraphics[width=0.26cm]{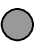}
\end{center}\end{minipage} &
\begin{minipage}[c]{13cm}
\raggedright
$(1-\phi_s \phi_b)^3+ 2\phi_s \phi_b(1-\phi_s\phi_b)(1-\phi_s
\phi_b(2-\phi_b))z_1+\phi_s^2\phi_b^2(1-3\phi_s\phi_b + 3\phi_s\phi_b^2
-\phi_s\phi_b^3)z_1^2+\phi_s\phi_b(1-\phi_s \phi_b(2-\phi_b))^2z_2 +
2\phi_s^2 \phi_b^2(3-2\phi_b)(1-\phi_s \phi_b(2-\phi_b))z_1z_2+\phi_s^3
\phi_b^3(8-11\phi_b+4\phi_b^2)z_1^2 z_2$
\end{minipage} \\
& & \\ \hline
& & \\
\begin{minipage}[c]{1cm}\includegraphics[width=1cm]{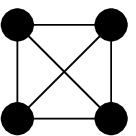}
\end{minipage} &
\begin{minipage}[c]{1cm}\includegraphics[width=0.26cm]{FilledDot.eps}
\end{minipage} &
\begin{minipage}[c]{13cm}
\raggedright
$(1-\phi_s\phi_b)^3 + 3\phi_s\phi_b(1-\phi_s\phi_b(2-\phi_b))^2z +
3\phi_s^2\phi_b^2(3-2\phi_b)(1-3\phi_s\phi_b+3\phi_s\phi_b^2-\phi_s\phi_b^3)z^2
+ \phi_s^3\phi_b^3(16-33\phi_b+24\phi_b^2-6\phi_b^3)z^3$
\end{minipage} \\
\end{tabular}
\caption{Subgraph generating functions~$F_r$ for all biconnected subgraphs
  of four vertices or fewer.}
\label{tab:examples}
\end{table*}

As an example, consider pure site percolation on a subgraph that takes the
form of an $m$-clique---a set of~$m$ vertices with an edge between every
pair.  The vertices in a clique have only one role, and an occupied
vertex~$i$ can reach another vertex~$j$ if and only if $j$ is itself
occupied.  Thus the distribution~$p_r$ is binomial in this case:
\begin{equation}
p_r(\phi;d) = \binom{m-1}{d} \phi^d (1-\phi)^{m-1-d},
\end{equation}
and
\begin{align}
F_r(\phi;z) &= \sum_{d=0}^{m-1} \binom{m-1}{d}
               \phi^d (1-\phi)^{m-1-d} z^d \nonumber\\
            &= (1 - \phi + \phi z)^{m-1},
\end{align}
where $\phi$ is the vertex occupation probability.

\begin{figure}
\begin{center}
\includegraphics[width=8cm]{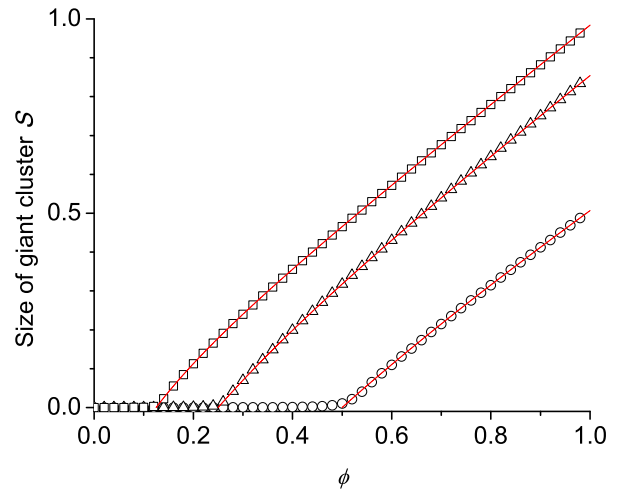}
\end{center}
\caption{The size~$S$ of the giant percolation cluster in the network of
  cliques described in the text, with cliques of size 2 to~5, as a function
  of the site percolation probability~$\phi$.  The circles, triangles, and
  squares show simulation results averaged over $100$ networks with $10^5$
  vertices each for $\av{k}=2$, 4, and~8 respectively, and the solid lines
  are the corresponding analytic results.  If role~$r=m$ is the role played
  by vertices in cliques of size $m$ then the average role indices are
  $\av{d_m} = \frac14\av{k}/(m-1)$ (which means that a randomly chosen edge
  is equally likely to belong to a clique of any size).}
\label{fig:graph2}
\end{figure}

Suppose, for instance, that we have a network made entirely of cliques of
various sizes~$m$ having one role each labeled with role labels~$r=m$, and
suppose the corresponding role indices~$d_m$ are independently Poisson
distributed at each vertex.  Then $G_m(\vec{d})=G_0(\vec{d})$ for all~$m$
and
\begin{equation}
u_m = [1 - \phi + \phi G_0(\vec{u})]^{m-1},\quad
S = \phi[1 - G_0(\vec{u})].
\end{equation}
Thus in this case we have $u_m = (1-S)^{m-1}$ for all~$m$ and $S$ satisfies
the self-consistent equation
\begin{align}
S &= \phi [ 1 - G_0(1-S,(1-S)^2,(1-S)^3,\ldots) ] \nonumber\\
  &= \phi [ 1 - H(1-S) ],
\label{eq:percs}
\end{align}
where $H(z)$ is the generating function for the ordinary degree
distribution, defined in Eq.~\eqref{eq:gfdegree}.  The percolation
transition in this network corresponds to the point where the gradient of
$\phi[1-H(1-S)]$ at~$S=0$ is~1, i.e.,~the point
\begin{equation}
\phi_c = {1\over H'(1)} = {1\over\av{k}}.
\end{equation}
Note that this is the same condition as for the percolation point on an
ordinary Poisson random graph, although the network is quite different.  In
Fig.~\ref{fig:graph2} we show a plot of $S$ from Eq.~\eqref{eq:percs},
along with the results of numerical simulations of finite sized networks in
this class.  As the figure shows, the agreement between simulation and
theory is excellent.

The computation of~$F_r$ for general subgraphs is typically more involved
than for the simple clique, since one must take into account possible paths
between vertices and allow for different roles.  Nonetheless, the
computations are in principle merely a matter of counting the number of
reachable vertices of each role for each possible configuration of occupied
vertices or edges and multiplying by the probability of the configuration.
We give in Table~\ref{tab:examples} the generating functions for all roles
of all biconnected subgraphs of four vertices or fewer (plus the single
edge), calculated by exhaustive enumeration in this fashion.  Generating
functions for singly connected subgraphs are simple composites of the
biconnected ones.

If we only wish to determine the position of the percolation transition,
the full subgraph generating functions are not necessary.  Linearizing
around the trivial fixed point in the equations for~$\vec{u}$ we again
derive an equation $\boldsymbol{\epsilon} = \mat{M}\boldsymbol{\epsilon}$,
and the existence of a giant component depends on whether the
matrix~$\mat{M}$ has any eigenvalues greater than one.  In this case
$\mat{M}$ is given by
\begin{equation}
M_{rs} = \sum_t \biggl[ {\partial F_r\over\partial z_t}
  {\partial G_t\over\partial z_s}\biggr]_{\vec{z}=\vec{1}}.
\end{equation}
The derivative $[{\partial F_r/\partial z_t}]_{\vec{z}=\vec{1}}$ is equal
to the mean number of vertices of role~$t$ reachable from a vertex of
role~$r$ within the appropriate subgraph.  It is an open question whether
there exists a method for calculating this mean number faster than the
exhaustive enumeration of states used to calculate the generating function
itself.

\section{Conclusion}
\label{sec:conclusion}
In this paper we have proposed and analyzed a random graph model that
incorporates an arbitrary distribution of any chosen set of subgraphs or
motifs, and hence mimics the properties of real-world networks, which are
observed in many cases to contain certain subgraphs in significant numbers.
The model is easily treated numerically and many of its properties can be
calculated analytically by virtue of a mapping to a locally tree-like
bipartite graph.  In particular, we have given calculations of subgraph
counts, the size of the giant component, the position of the transition at
which the giant component appears, and percolation properties for site and
bond percolation.

Useful though the model may be, however, it leaves open some important
questions.  We have not considered, for instance, how one should select the
set of subgraphs to be used.  If we wish to model a particular real-world
network, then we would presumably want to look at the subgraphs that appear
in that network and mimic their density and placement as accurately as
possible in the model.  But in that case, which subgraphs should be
considered to occur sufficiently frequently as to require their inclusion
in the model?  And what should the distribution of the various subgraphs
be?  For biconnected subgraphs the density in the model network is, as we
have shown, simply the density of the subgraphs we add explicitly, since
those created by the combination of other subgraphs have density zero.  For
singly connected subgraphs, however, the density is more complicated,
containing as it does not only the subgraphs we add ourselves but also
those made from other subgraphs, and at present we do not have a good way
to calculate it, making the matching of the real and model networks a
challenge.

Some further generalizations of the model are possible.  One could consider
subgraphs in which the stubs are labeled, for instance with colors, so that
even vertices in the same role could be distinguished.  By specifying the
colors of stubs connected to each vertex as well as the roles one could
then induce additional kinds of structure, such as bipartite or $k$-partite
structure or assortative mixing.  One could also include role--role
correlations, by analogy with the degree--degree correlations studied in
ordinary random graphs.

The treatment given in this paper also leaves open some mathematical
questions.  In particular, it is unclear whether there is a quicker way to
calculate the crucial subgraph generating functions of
Section~\ref{subsec:percolation} than by exhaustive enumeration of states.
For certain families of graphs, such as cliques, we have shown that it is
possible to characterize the generating functions analytically, but it
seems unlikely that this will be possible in more general cases.  It is
possible, however, that one could find a numerical algorithm for
calculating the coefficients of the generating functions more quickly than
the current method, which is exponentially slow.

\begin{acknowledgments}
  The authors thank Lada Adamic, James Gleeson, and Lenka Zdeborova for
  useful comments.  This work was funded in part by the National Science
  Foundation under grant DMS--0804778 and by the James S. McDonnell
  Foundation.
\end{acknowledgments}


\begin{thebibliography}{26}
\bibitem{note1} An exception is the geometric random
  graph~\cite{Penrose03}, in which clustering occurs naturally on account
  of the graph's being embedded in a Euclidean space of low dimension
  (usually two).  Geometric graphs are, however, something of a special
  case, most real networks having no such low-dimension embedding.
\bibitem{note2} One needs an additional constraint to avoid pathological
  graphs with two (or even more) giant components.  Consider an ensemble
  with two subgraphs, a cycle and a clique, and assume that the probability
  distribution is such that a vertex participates either in cycles or in
  cliques, but not both.  Then all graphs consist of a cycle part and a
  clique part that are disconnected from each other, so there could be two
  giant components, one composed of cycles and the other composed of
  cliques.  To avoid this pathology, we impose the restriction that one
  should not be able to divide the roles into two sets such that every
  vertex and every subgraph only has roles from one of the sets.
\end{thebibliography}
\end{document}